\documentclass[runningheads]{llncs}

\usepackage{graphicx}
\usepackage{latexsym}
\graphicspath{{./images/}}
\usepackage{booktabs} % for formal tables
\usepackage{color}  % for coloring text
\usepackage{amsmath}  % for aligning equations
\usepackage{subcaption}
\usepackage{caption}
\usepackage{tikz}
\usepackage{colortbl} % for color in tables
\usepackage{framed}
\usepackage{multirow}
\usepackage{multicol}
\usepackage{hyperref}
\usepackage{url}
\usepackage{balance}
\usepackage{verbatim}
\usepackage{cancel}

\newcommand{\squishlist}{
	\begin{list}{$\bullet$}
		{ \setlength{\itemsep}{0pt}
			\setlength{\parsep}{1pt}
			\setlength{\topsep}{1pt}
			\setlength{\partopsep}{0pt}
			\setlength{\leftmargin}{1.5em}
			\setlength{\labelwidth}{1em}
			\setlength{\labelsep}{0.5em} } }
	
	\newcommand{\squishend}{
\end{list}  }

\newcounter{qcounter}

\newcounter{tisctr}
\renewcommand{\thetisctr}{\arabic{tisctr}}

	{\normalsize \end{list} 
	\vskip 8pt}

	{\normalsize \end{list} 
	\vskip 8pt}

\begin{document}
	
\title{Towards Query Logs for Privacy Studies:\\
	On Deriving Search Queries from Questions}

\titlerunning{Towards Query Logs for Privacy Studies}
\author{Asia J. Biega\inst{1} \and
	Jana Schmidt\inst{2} \and
	Rishiraj Saha Roy\inst{2}
	\authorrunning{Biega et al.}
	\institute{Microsoft Research, Montreal, Canada
		\and Max Planck Institute for Informatics,
		Saarland Informatics Campus, Germany \\
		\email{asia.biega@microsoft.com, \{jschmidt, rishiraj\}@mpi-inf.mpg.de}}
}
	
	\maketitle
	
	\begin{abstract}
Translating verbose information needs into crisp search queries is 
a phenomenon that is ubiquitous but hardly understood. 
Insights into this process could be valuable in
several applications, including synthesizing large privacy-friendly query logs 
from public Web
sources
which are readily available to the
academic research community. 
In this work, we take a step towards understanding query formulation by tapping
into the rich potential of community question answering (CQA)
forums. 
Specifically, we sample
natural language (NL) questions spanning diverse themes from 
the Stack Exchange platform, and conduct a large-scale conversion experiment
where
crowdworkers submit search queries they would use when looking for equivalent
information. 
We provide a careful analysis of this data, accounting for possible sources of 
bias during conversion, 
along with
insights into user-specific
linguistic patterns and
search behaviors.
We release a dataset of $7,000$ question-query pairs from this study
to facilitate further research on query understanding.

\end{abstract}

	% !TEX root = ../2019-arxiv-stacklog.tex
\section{Introduction}
\label{sec:intro}

\textbf{Motivation.} Detailed query histories often contain a precise picture
of a person's life, including sensitive and personally identifiable information.
% Because
As sanitization of such logs 
%is a difficult problem that escapes easy solutions such as k-anonymity,
is an unsolved research problem,
% search companies
commercial Web search engines
% who have
that possess large datasets of this kind at their disposal refrain from
disseminating
% search logs
them to the wider research community. 
These concerns were made obvious especially after the 2006 AOL log
scandal\footnote{\url{https://en.wikipedia.org/wiki/AOL_search_data_leak},
	Accessed 06 Jun 2019.}.
Ironically, however, studies examining privacy in search often require
detailed search logs with user profiles
\cite{biega2017solidarity,biega2016r,adar2007user,chen2011ups,zhang2016safelog,zhu2010anonymizing,xu2007privacy,jones2007know}.
Even beyond privacy, collections with rich interaction profiles of users
are also an asset in personalization~\cite{bennett:12,ustinovskiy:15,wang:13}
and simulation
studies~\cite{huurnink2010simulating,azzopardi2011report,berendsen2013pseudo} 
in information retrieval.
% berendsen2013pseudo}.

% \noindent \textbf{Limitations of current resources.} 
While there exist a number
of public IR collections, none of them contain data necessary for such studies.
Notable among these,
% praise trec a bit
the TREC Sessions Track 2014 data~\cite{carterette:14} has $148$ users,
$4.5k$ queries and about $17k$ relevance judgments. 
There are roughly ten sessions per user, where each session is usually a set
of reformulations.
% A collection like this
Such collections
% constitutes a 
are rather small %-scale dataset
for
driving research in user-centric privacy.
% the purpose of studying user privacy.
The 2014 Yandex collection released as part of a
workshop on log-based personalization~\cite{serdyukov:14} is useful for
evaluating personalization algorithms~\cite{ustinovskiy:15,cai:14}. 
However, to protect the privacy of Yandex users, every query term is replaced
by a meaningless numeric ID.
%\footnote{\url{
% https://www.kaggle.com/c/yandex-personalized-web-search-challenge/data},
% Accessed 05 November 2018}. 
This anonymization strategy makes semantic interpretation impossible and
may be a reason why
this resource did not receive widespread adoption in privacy studies.
Interpretability of log contents
% is often
is important to understand privacy
threats
\cite{biega2016r,biega2017solidarity,jones2007know,chen2011ups,xu2007privacy}.
%for personalization over long-term profiles.

% \todo{Perhaps talk about simulation in IR as another imperfect alternative.}
%\todo{Talk about other types of collections people have derived
% from Stack Exchange.}

Motivated by the lack of publicly available query logs with rich user profiles, 
Biega et al.~\cite{biega2017solidarity} synthesized a query log from
the Stack Exchange\footnote{\url{https://stackexchange.com/sites},
Accessed 06 Jun 2019.}
% rishi: probably the sites conveys the diversity aspect better?
platform
 -- a collection of CQA subforums on a
 multitude % variety
 of topics.
 Each subforum is focused on a specific topic, like % ranging from
 % including
 % mathematics % rishi: we remove maths and code later
 linguistics, parenting, and beer brewing.
 % and many more.
% rishi: make the correspondence between subforums and topics
% 1 subforum, 1 topic
Queries in the synthetic log were derived from
users' information needs posed as 
NL questions. %, which are a natural way of describing information 
%needs.
A collection like this has three advantages. First, it enables creation of
rich user profiles
by stitching queries derived from questions asked by the same user across
different topical forums.
Second, since it was derived from public resources created by users under 
the Stack Exchange terms of service 
(allowing for reuse of data for research purposes), 
it escapes the ethical pitfalls intrinsic to dissemination of private user data.
Third, CQA forums contain questions and
assessments of relevance in the form of accepted answers \textit{from the
	same user}, which is
% unfortunately not the case in most IR forums like TREC,
% but is absolutely 
vital for the correct interpretation of
query intent~\cite{chouldechova2013differences,bailey:08}.
The proposed derivation approach in~\cite{biega2017solidarity},
however, was rather heuristic: the top-$l$
TF-IDF weighted question words were extracted to form a keyword query, where
the query length $l$ was uniformly sampled from a range of one to five words.

% rishi: appears a bit detailed
%The proposed derivation strategy, however, used a simple heuristic for
% converting natural language questions to queries --
%a length $l$ of the query between 1 and 5 was sampled from
% a uniform distribution,
%and the top-$l$ question words with highest TF-IDF were chosen for form
% the query.

\textbf{Contributions.}
% Recognizing that
Such a query log derivation
methodology from CQA forums
has the potential to
% deliver
produce sizeable IR collections,
a fact recognized by recent analogous efforts from
Wikipedia~\cite{sasaki2018cross}.
%for several studies,
% on topics such as search privacy or personalization,
% repeated often
% So a natural follow-up is to make derived queries more realistic.
% an important follow-up step is to ensure a tighter resemblance with 
% non-synthetic querylogs.
% rishi: looks roundabout
However, to harness CQA resources better, % To this end,
it is necessary to: (i) better 
understand how humans formulate keyword queries
given a verbose information need in natural language,
% for specific information needs 
% represented by natural language questions,
and, (ii) derive other %% IR collection
elements like
candidate documents and relevance judgments
from CQA forums for completeness of derived benchmarks.
% These problems are the main focus of this paper.
This paper focuses on these problems and makes the following contributions:
\squishlist
% Our main contribution is
\item We conduct a large-scale user study 
where crowdworkers
% are asked to
convert questions % describing information needs
to queries;
\item We provide insights from the collected data that can drive strategies
for automatic conversion at scale;
% A careful analysis of the data, accounting for possible sources of bias during 
% conversion,
% provides insights into linguistic and demographic patterns in query 
% formulation.
\item We release $7,000$ question-query pairs % collected in this study
that can be used for training and evaluating such conversion methods at
\url{https://www.mpi-inf.mpg.de/departments/databases-and-information-systems/research/impact/mediator-accounts/};
% for further research.
\item
% As an auxiliary contribution
We propose a methodology for deriving other collection elements 
like
 documents and relevance labels
from
% the content and structure of
CQA forums and analyze the utility of such a % collection. % twice
resource.
% and empirically analyze the characteristics and
%\item We analyze the % characteristics and
%utility of such a 
%% Stack Exchange-derived IR collection.
%derived resource in IR tasks.
\squishend

\section{User study: Setup and Pilot}
\label{sec:user-study}

% 
% In this work, we seek to understand how humans construct queries from
% information needs to form
% a ground for better conversion methods. 

% The goal of the
We conduct a large-scale user study % is
to
% find out
understand how humans create keyword queries for
% specific
information needs
% described by
expressed as natural language questions. 
We use questions from the % publicly available
Stack Exchange forum
and ask workers on Amazon Mechanical Turk (AMT) to create queries
specifying
% describing
the same
information needs as the questions.
Notably, crowdsourcing has recently been successfully applied
to similar creative tasks
of short text generation~\cite{chen2018user},
where workers paraphrased search result snippets.

% This section describes the study in detail.

%\abcomment{Perhaps mention the difference to query reduction studies.}
% rishi: is it really necessary here?

% !TEX root = ../2019-arxiv-stacklog.tex
\subsection{Question sampling}
\label{subsec:qsamp}

%\textbf{Choosing subforums.} 
\textbf{Filtering subforums.} 
% The first step in the study is to collect questions 
% for conversion that meet certain desiderata. 
We used the Stack Exchange dump from March
2018\footnote{\url{https://bit.ly/2JI8ubn},
Accessed 06 Jun 2019.}
with data for more than $150$ different subforums.
We are interested in questions written in \textit{English text}
and thus exclude forums primarily dealing with programming, mathematics,
and other % natural
languages like French and Japanese.
% (for example, the subforums ``Code Review'', 
% ``MathOverflow'', and``French'') were removed from further consideration.
% These left us with $140$ subforums to choose questions from.
Moreover, we want to avoid highly-specialized forums
as the average AMT user
may not have any background knowledge to generate queries for such niche 
domains. 
% Next, we did not to have questions from tail forums
% that reflect highly specialized interests (for example, ``Augur'', a 
% decentralized oracle and prediction market platform) as the average AMT user
% may not have any background knowledge to generate queries from such niche 
% domains. 
To this end, we exclude all subforums with less than
$100$ questions, as a proxy for expression of a critical mass of interest.
We found that $75$ subforums satisfy our requirements for
subsequent question sampling.

\textbf{Filtering questions.} 
% Since the end goal of this study is to derive
% an IR benchmark,
% gold standard relevance assessments for sampled questions are indispensable.
% As mentioned earlier, 
% these are encoded as ``Accepted answers'' by askers on StackExchange.
% So as the next step,
% we removed all questions that did not have answers accepted by
% the original asker. A substantial $52\%$ of the questions in these
% $75$ subforums were found to have an accepted answer (a positive relevance
% label), testifying the general success of CQA forums like StackExchange
% or Quora. To make the resultant collection interesting from a document ranking
% perspective, it is important that a query (originating from a question) have
% more than one candidate pseudorelevant document. Note that answers to
% a question
% form its document pool, or the question-relevant corpus, in our collection.
% As a result, we only sample questions that have at least five answers provided
% by the community.
As a proxy for questions being understandable by users,
we choose only those that have an answer accepted by the question author,
and with at least five other answers provided.

\textbf{Sampling questions.}
Under these constraints, we first sample $50$ subforums from the 
$75$ acceptable ones to have high diversity in question topics.
Next, we draw $100$ questions from each of these
thematic groups, producing a sample of $5000$ questions,
which is used as the data in the user study.

\subsection{Study Design}
\label{subsubsec:study-design}

AMT crowdworkers in the study were first presented with a short tutorial
explaining the task
as well as a number of examples visualizing the task. 
After familiarizing themselves with the instructions, they proceeded to 
converting questions to search queries in the main part of the study,
followed by a survey on demographic information.

\subsubsection{AMT Setup}
\label{subsubsec:amt-setup}

We recruited a total of $100$ workers who had Master qualifications
and an approval rate of over $95\%$ to ensure quality of annotations. 
We paid $\$6$ per assignment in the pilot study ($30$ questions per assignment),
and $\$9$ per assignment in the main study ($50$ questions per assignment.)
The workers were given three hours to complete the assignments, while
the actual average time taken per question
turned out to be $116$ seconds.
%Table~\ref{tab:studygeneraloverview} presents an overview of the
%main AMT study (Sec.~\ref{subsec:main-study}).
% 
%\begin{table}[h]
%	\centering
%	\begin{tabular}{l l l}
%		\toprule
%		\textbf{Property} 	& & \textbf{Value}							\\ \toprule
%		Title				& &	Write a Web search query					\\
%		Description			& & Given a forum post formulate a single query	\\
%		Keywords			& & Forums, Question answering, Queries			\\
%		Reward				& & \$9 per HIT									\\
%		Requirements		& & Workers must be Masters						\\
%		Time allotted		& & Three hours									\\
%		Time taken			& & $1.6$ hours 								\\ \bottomrule
%		% Total cost 		& & 1537 EUR  									\\ \bottomrule
%	\end{tabular}
%	\caption{Overview of the main AMT study.}
%	\label{tab:studygeneraloverview}
%\end{table}

\subsubsection{Instructions and Task}
\label{subsubsec:instructions}

\begin{figure}[t]
	\includegraphics[width=\columnwidth]
	{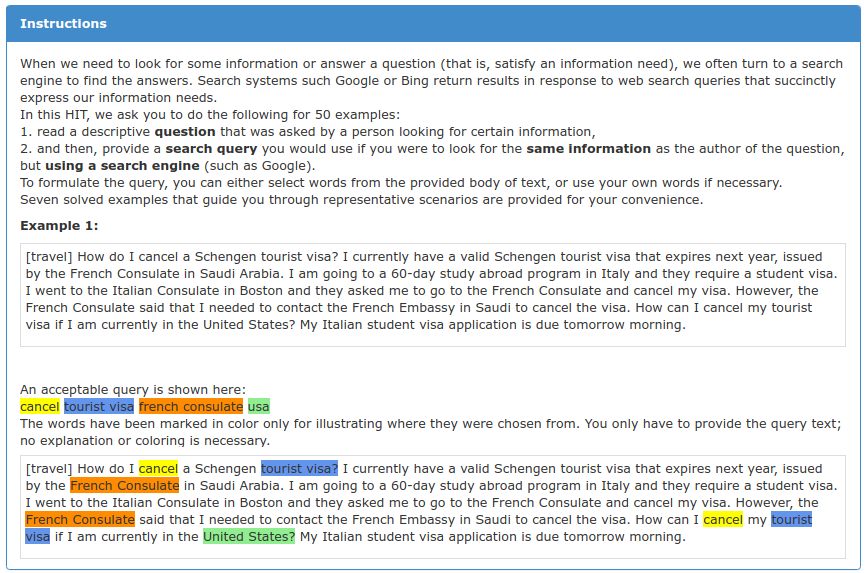}
% 	{instruction-screenshot-pilot-highres.png}
	\caption{Instructions from the pilot study.}
	\label{fig:instruction-screenshot-pilot}
	% \vspace*{-0.7cm}
\end{figure}

We kept the task guidelines simple to avoid biasing participants towards
certain answers.
We do mention the workers are free to select words from the text of
the question 
or use their own words. Fig.~\ref{fig:instruction-screenshot-pilot} shows
a screenshot with the instructions
from the pilot study. The instructions give a very high-level overview of
the procedure of
searching  for information using search engines to introduce the context.

We provided several examples as a part of the instructions to better
illustrate the task.
Examples were meant to cover the various ways of arriving at
a correct solution:
selecting words from the question, using own words, changing grammatical forms, 
constructing queries of various length, etc.
These cases were not made explicit, but communicated by highlighting words
in the text
and the query, as shown in Fig.~\ref{fig:instruction-screenshot-pilot}.

As a main task, participants were shown a number of questions expressing
certain information needs
and asked to formulate search queries they would use 
to search for the same information as the author of the question was
searching for. The questions were presented in the following form:\\
 \centerline{ $[$Forum Name$]$ Title Body}\\
Each question was a concatenation of the Stack Exchange post title and body,
and prefixed with the forum name the post comes from to give it
the right context.
% \subsubsection{Demographic survey}
The main task was further accompanied by an optional demographic survey
to help us understand if various demographic features influence
how people formulate queries.

% !TEX root = ../2019-arxiv-stacklog.tex
\subsection{Pilot study}
\label{subsec:pilot}

We tested the setup in a pilot study with five HITs and $30$ questions each.
The average query length was $5.65$ words with a standard deviation of $2.40$. 
Out of $150$ questions, the forum name was included in the corresponding query
$33$ times.
In nine of such cases, the forum name was not present in the title or body of
the question,
which suggests the presence of the forum name is important in determining
the context of the question.
Most query words were chosen from the title,
although title words are often repeated in the body of the question.
Workers used their own words or words modified from the question $47$ times.
These results suggest that participants generally understood the instructions.
The five workers took $22$, $75$, $92$, $88$, and $100$ minutes to complete
the tasks.
%One of the workers was much faster than others: the demographic survey
%revealed that the they work in the IT
%industry and thus might be familiar with search engines and query formulation.
The demographic survey revealed that the fastest person with $22$ minutes
worked in the IT industry.

% !TEX root = ../2019-arxiv-stacklog.tex

\section{User study: Controls and Main}
\label{sec:user-study-main}

\subsection{Control study for title position bias}
\label{subsec:bias}

A vital component of any crowdsourced study is to check if participants are
looking for quick workarounds for assigned tasks that would make it hard
for requesters to reject payments, and to
control for confounding biases. In the current study for converting questions
to queries, the source of potential bias stems from the fact that a question
is not just
a sequence of words but a semi-structured concept: it has (i) a subforum name
to which the question belongs,
(ii) a title, and (iii) a body. Web users might be aware
that question titles in CQA forums often summarize the question. Thus, if the 
structure is apparent to the annotator, she might use words
only from the title without meticulously going through the full question content,
which may often be a few hundred words.

To mitigate such a possibility, we present the titles in the same
font 
%(not magnified, bold, italicized,  or colored) 
as the body, and do not separate them with
% unnecessary whitespace.
newlines.
%(such as a blank line)
% and present
%them as one coherent paragraph. 
Regardless of this uniformity of presentation,
%so that the title no longer stands out, 
users may still be able to easily figure out
that the first sentence is indeed the question title. To quantify the effect
of this \textit{position bias} of the title, we used ten HITs ($500$ questions)
as a \textit{control experiment} where, unknown to the Turkers, the title was
appended as the \textit{last sentence} in the question. These same
$500$ questions were also annotated in the usual setup in the main part of
the study for comparison.
 
We compare the results of the main and the control studies by
measuring the fraction of times users chose words from the first and
the last sentences. 
Results are shown in Table~\ref{tab:control}.
%which report the fractions of words chosen by users from
%these question zones (expressed as percentages) to feature in the query,
%over these $500$ questions.
% Presence of ``exclusive'' counts cases when selected words did not appear
% elsewhere in the question, like the body text or the subforum name. 
% Statistical significance tests were performed with the $2$-tailed paired
% $t$-test, and the null hypothesis was rejected if $p < 0.05$.
%The percentages
%for titles were computed after collecting annotations; Turkers did not know
%at the time of the study that these were titles. Note again that in
%the main study,
%titles were the first sentences, while they were put at the end in the control
%experiment.
Fractions were normalized by the length of the question title, as raw counts
could mislead the analysis (longer question titles automatically contribute
larger \textit{numbers} of words to the queries). 
Thus, if two words were
chosen from an eight-word-long title to appear in the query, we would compute
the reported fraction as $0.25$ $(25\%)$.

We make the following observations: (i) in both the main study
and the control, users choose words from the title very often 
($\simeq 97\%$ and $\simeq 84\%$, respectively), showing similar
task interpretation.
Note that such high percentages are acceptable, as question titles typically
do try to summarize
intent. Nevertheless, examining the entire body for complete
understanding of the asker's intent is central to our task.
(ii) Relatively similar percentages of query words originate from the
titles in both cases ($37.7\%$ vs. $26.1\%$).
(iii) If Turkers were trying to do the task
just after skimming the first sentence (which they would perceive as 
the title), the percentage of words from the first sentence in the
control would have been far higher than a paltry $12.2\%$, and the last sentence
would contribute much lower than a promising $26.1\%$.

Words in
the title, being topical to the question, are often repeated in the body text in 
CQA forums. To test this word selection owing to presence elsewhere in
the question
(like body text or subforum name), we also measured the fraction of words
chosen by Turkers in the control study that occurred \textit{exclusively} in
the last sentence.
This was found to be $4.1\%$: we posit that if Turkers were simply looking
to make
quick money, or biased by imagining the first line to be the title,
this fraction
would have been close to zero. We thus conclude that Turkers completed their
HITs with due understanding and sincerity: words from titles were chosen
frequently because of their \textit{relevance}, and not
because of their relative positions in the question. As a result of
this study, we chose to
keep titles in their original first positions for the remainder of
the main study, as
putting it at the end degrades the overall coherence and readability of
the question.

\begin{table} [t]
	% \small	
	\centering
	\begin{tabular}{l c c}
	\toprule
		\textbf{Property}							& \textbf{Main study}		& \textbf{Control study}	\\ \toprule	
		Times question title word chosen for query	& $96.6\%$					& $83.8\%$					\\
		Question title words in query				& $37.7\%$					& $26.1\%$					\\	\midrule		
		% Title (Exclusive) 						& $7.7\%$					& $4.1\%$					\\			
		Question first sentence words in query		& $37.7\%$					& $12.2\%$					\\			
		% First sentence (Exclusive)				& $7.7\%$					& $3.0\%$					\\			
		Question last sentence words in query		& $9.0\%$					& $26.1\%$					\\ \bottomrule	
		% Last sentence (Exclusive) 				& $1.4\%$					& $4.1\%$					\\ \bottomrule
		% rishi: p-value column in confusing; exclusive is also confusing - what if same words were chosen due to
		% presence in body and not in title 
	\end{tabular}
	\caption{Measurements from the position bias control study.}
%	\caption{Measurements from control study showing normalized percentages
% of query words chosen from various parts of the question text.
%	% ``Exclusive'' indicates words were not present elsewhere in the question,
% like body or subforum name.
%%	Values substantiate
%%	that Turkers performed the task sincerely, and without being biased by
% position of the title.
%	}
  \label{tab:control}
  % \vspace*{-1.0cm}
\end{table}

% !TEX root = ../2019-arxiv-stacklog.tex
\subsection{Control study for user agreement}
\label{subsec:control-user}

While the main focus of the study was to construct a sizable collection
of question-query pairs, we were also interested in learning 
how robust query formulation is to individual differences.
To this end, we issued a batch of 10 tasks with 50 questions to be
completed by three workers each. The validity of the comparison comes from the 
experimental design where query construction is conditioned on
a specific information need.
% \abcomment{Perhaps add some citations to other papers comparing
% queries between users
% but without such conditioning}.

% Table \ref{tab:multipleuser} presents the term overlap statistics.
% Overlap between pairs is computed  by XXX.
% Overlaps between all three is computed by YYY.
% We also normalize the value by dividing the overlap by
% the length of the longest query.
We compute the average Jaccard similarity coefficient between all pairs of
queries $(q_1, q_2)$
corresponding to the same question:
$Jaccard(q_1, q_2) = \frac{|q_1^W \cap q_2^W|}{|q_1^W \cup q_2^W|}$,
where $q_1^W$ and $q_2^W$ are
the sets of words
of the compared queries. We find the average overlap to be $0.325$,
and to come mostly for the most informative content words in the question.

\subsection{Main study}
\label{subsec:main-study}

\subsubsection{Data Collection}
\label{subsubsec:collection}

The main study was conducted with insights obtained from the pilot and
the control studies. In total, we asked $100$ AMT users to convert $5000$ 
questions to queries.
Users who participated in control studies were
not allowed to take part again, to avoid \textit{familiarity biases}
arising from
such overlap. 
Basic properties of this stage are presented in
Table~\ref{tab:main}. 
% The title,
%description, and keywords were intentionally kept very simple to make the task
%visible to a large number of Turkers. 
Guidelines were kept the same as
in the pilot study.
%, as they seemed to work quite well. 
%The title was blended into the
%question body to preclude word selection biases arising from knowledge that
%a given sentence is the question title (Sec.~\ref{subsec:bias}). An additional
%question on the number of search queries issued per day by the Turker
%was added to
%the demographic survey (Sec.~\ref{subsec:demographics}).
%rishi: pointer to next subsection here
%Such activity could be correlated with search expertise,
%and such expertise may manifest itself subtly in the style of queries generated.
The number of questions in one
HIT were increased from $30$ to $50$, to cover
user-specific querying traits better. 
In line with this change, the reward per HIT was
increased from $\$6$ to $\$9$. 
%The only restrictions that we imposed
%were that: (i) the Turker must possess the Master qualification on AMT
%($\geq 95\%$ success rate on past tasks), for an expectation of cleaner
%results, and (ii) one Turker cannot do multiple
%% HITs, as that would defeat the intent to collect data from as many distinct
%users as possible (a target of $50$ in this case). 
The Turkers took about
$1.6$ hours per HIT, which comes down to $116$ seconds per question. 
Since Stack Exchange questions can be quite long, we
believe that such a mean task completion time is reasonable. The mean
query length turned out to be $6.15$ words,
which is longer than the
average Web search query (about
three to four words~\cite{saharoy2014unsupervised,yom2018demographic}).
We believe that this is likely because Stack Exchange questions 
express more complex information needs.

\begin{table} [t]
	\centering
	\begin{tabular}{p{4cm} p{8cm}}
	\toprule
		\textbf{Property}	& \textbf{Value}								\\ \toprule	
		Title 				& ``Write a Web search query''					\\
		Description 		& ``Given a post, formulate a single query.''	\\					
		Keywords			& Question answering, Queries, Web forums		\\
		Questions in a HIT	& $50$											\\
		Total HITs			& $100$											\\
		Reward				& $\$9$ per HIT									\\
		Time allotted		& $3$ hours per HIT								\\
		Time required		& $1.6$ hours per HIT on average				\\	
		Restrictions		& Workers must be Masters, One Turker one HIT 	\\ \bottomrule	
	\end{tabular}
	\caption{Summary of the AMT main study.}
  \label{tab:main}
\end{table}

%The table below describes the general settings for this study as well as
%the total payment. This section has a similar structure as the pilot study but 
%some adjustments
%were necessary. First of all the changes explained above were applied.
% Furthermore,
%the number of questions per task was increased from 30 to 50. So one user
% needs to
%answer more questions which results in a better user specific analysis because 
%more data
%from one user is present. Overall more than 4500 queries were generated
%in this study
%part. The questions are chosen again randomly and come from different forums
%to get
%diversity of topics for one participant.
%
%\abcomment{Table missing.}
%

\subsubsection{Data Analysis}
\label{subsubsec:analysis}

%The goal of the study is not just to collect 
%user-generated queries for a benchmark, 
%but rather to understand how humans formulated queries from
% descriptive information needs. 
%%To this end,
%%we made some measurements on the $5000$ queries that we collected, to gain
%%insights into the query generation process. Such insights can be
%%subsequently encoded into developing faithful generative models for queries,
%%enabling conversion of questions at scale to derive rich IR benchmarks
%%from CQA forums. 
%Specifically, 
We looked into three aspects of 
\textit{question-query
pairs} when trying to discriminate between words that are \textit{selected}
for querying, and those that are not.

\noindent \textbf{Position.} 
%When a human formulates an information need in her 
%mind, it is manifested necessarily as a \textit{sequence} of words. Thus, e
%posit that 
%We word position plays a role in the keyword selection process.
We measured relative positions of query and non-query words in the question,
and found that a major chunk ($\simeq60\%$) of query words originate
from the first $10\%$ of the question. The next $10\%$ of the question
contributes an additional $17\%$ of words to the query; the remaining $80\%$
of the question, in a gently diminishing manner, produce the rest $13\%$
of the
query. This is a typical top-heavy distribution, suggesting humans 
conceptualize the core \textit{content} of the information need first 
%(which definitely
%go into the query), 
and gradually add specifications or conditions of
\textit{intent}~\cite{saharoy2014unsupervised} towards the end~\cite{saharoy:15,saharoy:13a}.
Notably, even the last
$10\%$ of the question contains $2.78\%$ of the query, suggesting that
we cannot disregard tail ends of questions.

Finally, note
that the title is positioned at the beginning of the question
(Sec.~\ref{subsec:bias}), and alone accounts for $57\%$ of the query.
Title words, however,
do repeat in the body, and further inspection reveals that $12\%$ of
the query mass is comprised of words that appear exclusively in the title. 
In Stack Exchange, title are often constructed as summaries of the questions.
%The intuition here
%can perhaps be more tangible if the title can be conceived as
%an initial summary of the question by the user.

We also allowed users to
use their \textit{own words} in the queries.
Our analysis reveals that a substantial $17\%$ of
query words fell
into this category. Such aspects of this data pose interesting
research challenges
for query generative models.
%, and will prefer methods that rely on
%language understanding, rather than simply keyword selection via
% term weighting.

\noindent \textbf{Part-of-speech (POS).} Words play roles of varying
importance in sentences, with a high-level distinction between \textit{content
words} (carrying the \textit{core} information in a sentence) and
\textit{function words} (specifying \textit{relationships} between content 
words).
%At a
%finer granularity, we have analogous parts-of-speech (POS): nouns, verbs, 
%adjectives,
%and so on.
Search engine users have a mental model of what current search engines
can handle: most
people believe that function words (prepositions, conjunctions,
etc.) are of little importance in query formulation, and tend to drop
them when issuing queries. These intuitions are indeed substantiated by our 
measurements: content words (nouns, verbs, adjectives and adverbs)
account for a total of $79\%$ ($47\%, 15\%, 13\%$, and $4\%$ respectively)
of query words, while function words constitute only $21\%$ of the query.
In this work,
we use the $12$ Universal POS tags (UTS)\footnote
{\url{https://github.com/slavpetrov/universal-pos-tags},
Accessed 06 Jun 2019.} proposed by Petrov et al.~\cite{petrov2012universal} .
%and subsequently shown to be useful in a wide
%array of NLP applications;
%the PTB
%tagset\footnote{\url{
%https://www.ling.upenn.edu/courses/Fall_2003/ling001/penn_treebank_pos.html}
%, Accessed 05 November 2018.} with $38$ tags is too fine-grained
% for our purpose 
%where we
%are trying to mimic the human thought process of judging word importance.
Our findings partially concur with POS analysis of Yahoo! search queries
from a decade back~\cite{barr2008linguistic} where nouns and adjectives
were observed to be the two most dominant tags; verbs featured in the seventh
position with $2.4\%$. We believe that the differences can be attributed to the 
changing nature of search, where more complex information needs demand more
content words to be present in the queries. 
%Also, note that the
%previous study did not have access to questions or descriptions, and thus
%directly applied POS taggers on queries, which is known to be error-prone.
%as tagging algorithms are trained on large volumes of natural language 
%corpora, from which query syntax is known to deviate considerably. 
These
insights from POS analysis of queries can be applied to several tasks,
like query segmentation~\cite{saharoy:14b,saharoy2014unsupervised,hagen2012towards}.

\noindent \textbf{Frequency.} A verbose information need may be characterized
by certain recurring units, which prompted us to measure the normalized
frequency $TF_{norm}$ of a term $t$ in a question $Q$, as
$TF_{norm}(t, Q) = TF(t, Q) / len(Q)$, where $len(Q)$ is the question length
in words. Query words were found to have a mean $TF_{norm}$ of $0.032$,
significantly higher than that of non-query words ($0.018$).

%\noindent \textbf{Summary.} We conclude that word position, part-of-speech,
% and
%frequency, all play a vital role in determining the likelihood of
% a question word
%being selected during query formulation.

% !TEX root = ../2019-arxiv-stacklog.tex
\subsection{Demographics}
\label{subsec:demographics}

% Demographics: At the end the users were asked to fulfill some
% demographic questions.
% After the pilot study, another highly interesting question about
% the daily Web search
% behavior was added. The Pearson correlation coefficient is used to determine
% if there
% exist a correlation between different attributes and the query length.

Workers in the study were asked to fill a demographic survey at the end of
the task.
We made these questions optional so as not to incentivize
fake answers if the workers feel uncomfortable giving an answer.
We asked about gender, age, country of origin, 
highest educational degree earned, profession, income,
and the frequency of using search engines as the number of search queries
issued per
day (such activity could be correlated with search expertise, and the
expertise may manifest itself subtly in the style of the generated queries).

From the $100$ subjects in our study, $50$ were female and $50$ were male.
Nearly all lived in the United States except for three who lived in India.
% We calculated the Pearson coefficient to understand 
% the correlation between different demographic attributes and the query length.
% We generally found no correlation between the query length and
% any of the features except for a week correlation with age and a very weak 
% correlation with gender.
We found a weak correlation between the query length and age (query length
generally increased with the age of participants),
anf found that men formed slightly longer queries on average ($6.56$ words
for men versus $6.15$ words for women).

\subsection{Released Data}
\label{subsec:rel-data}

We release a dataset of $7,000$ natural language questions paired with
corresponding search queries ($5,000$ from the main study
and $2,000$ from the control studies).
The average query length is over six words, reflecting a degree of complexity
in the underlying 
information needs, and in turn, interesting research challenges for methods
aiming 
at automated conversion strategies for synthetic query log derivation. 
Key features of this collection include: 
(i) question topics spanning $50$ different 
subforums of Stack Exchange, and (ii) question-query pairs grouped by
$100$ annotator IDs, making the released testbed suitable for analyzing 
user-specific query formulation, and cross-domain experiments.

	% !TEX root = ../2019-arxiv-stacklog.tex
% \section{Examining the potential of an IR collection derived
% from StackExchange}
\section{Potential of Derived IR Collections}
\label{sec:potential}

The insights from our user study could be used to drive automatic conversion
methods for synthetic query logs.
The collection derived by Biega et al.~\cite{biega2017solidarity}
contains just a query log with user profiles. 
However, many more elements of IR collections, including the notion of
document relevance, are embedded in the contents and structure of
CQA forums.
While NL questions represent information needs and can be converted to queries, 
answers to these questions are analogous to documents that satisfy these needs.
Moreover, systems of rating answers such as upvotes or acceptances by
the question author, are in fact explicit assessments of relevance.
%, crucial to IR evaluation. 
% explicitly express document relevance for a given information need.
In this section, we aim to analyze the characteristics and potential of such
a synthetic IR collection derived from the Stack Exchange forum.
%describe our source dataset and the method used to create our collection.
% in collection extraction method in detail.

\subsection{Deriving a collection from Stack Exchange}
\label{sec:method}

\subsubsection{Source dataset}
\label{subsec:stack}

We extract the collection from the Stack Exchange dump
%dated
% 13 June 2016\footnote{\url{https://archive.org/details/stackexchange},
% Accessed 27 February 2017}, % (dump as on 2016-06-13),
from March 2018.
% which was released under the Creative Commons license and is
% freely available for download.
It contains all information publicly available on any of
the $152$ thematically diverse subforums within
Stack Exchange.
\textit{Topics} range from the general domain, like fitness, beer brewing,
and parenting, to more technical areas, such as astronomy, engineering, or
computer programming. 
% Despite this topical diversity, technical subforums like Stack Overflow
% are still the major contributors to the overall content of Stack Exchange.
%contribute the most in terms of the dataset size.
We exclude the largest subforum Stack Overflow from the source dataset
to avoid the dominance of programming queries in our collection. 

Each subforum dump contains, among other content, all posted questions,
answers, and comments, as well as information about accepted answers,
upvotes and downvotes. % User profiles can be assembled from across
% subforums using the \emph{globalID} attribute,
%allowing for a full reconstruction of the interaction history of any user
% who has not deleted her profile. 
User \textit{profiles} can be constructed by joining questions and answers
with the same \textit{globalID} attribute across subforums. 
Such users profiles, often unavailable in other published IR collections,
can potentially be an asset for personalization algorithms.
%as they may help estimate rich user-level language models, among other things.

%\begin{table} [t]
%	\centering
%	\caption{Statistics of source Stack Exchange dataset.}
%	%\newcolumntype{G}{>{\columncolor [gray] {0.90}}c}
%	%\resizebox{\columnwidth}{!}{
%		\begin{tabular}{l r | l r }
%			%\begin{tabular}{c G G c c}
%			\toprule
%			\#Users		& 0000	& \#Questions	& 5555	\\ \bottomrule
%			\#Forums	& 5555	& \#Answers		& 5555	\\ \bottomrule
%		\end{tabular} %}
%	\label{tab:st_stats}
%\end{table}
%
%Table~\ref{tab:st_stats} presents statistics regarding the source dataset.

\subsubsection{Questions to queries}
\label{subec:log}

The results of our user study suggest that the term frequency features 
are indeed a reasonable indicator of whether a term should be included in
a query.
We thus follow the general methodology of
Biega et al.~\cite{biega2017solidarity} for converting queries to questions
where we choose a random number of question terms with the highest TF-IDF.
However, we modify the distribution from which the query lengths are sampled
to resemble the distribution 
estimated from the 2006 AOL log~\cite{pass2006picture}. 
The term frequency (TF) is measured within the question, and
the inverse document frequency (IDF) is calculated from the set of
all questions and answers.
We retain users with at least $100$ queries.

\noindent{\textbf{Duplicate queries.}} A key difference between query logs
and questions from strongly moderated forums like Stack Exchange
is the lack of duplicate information needs. % in different user profiles.
While a search engine log has many instances of repeated queries, in
a CQA forum a question is often closed, merged or deleted if
a similar question has been asked before.
%That is, questions often get closed, merged or deleted because
% another similar question has already been asked. While this difference
% is not as heavily visible when only subsets of keywords from
% the queries are extracted, we nevertheless
Thus, to simulate duplicate queries, we use another feature of Stack Exchange:
% to alleviate this problem,
\emph{marking a question as a favorite}. When a user marks
another user's question as a ``favorite'', they start following
the question and get notified about its updates. 
We interpret this as a signal of the user expressing the same information need. 
We thus duplicate a question in the histories of all users who marked it
as her favorite, \textit{before} the query extraction process.

\subsubsection{Answers to documents}
\label{subsec:docs}

Answers in Stack Exchange and other CQA forums like Quora are often several
paragraphs long, and can naturally be treated as documents for an IR collection.
% from the source Q\&A forum can naturally serve as an underlying document
% collection for testing the retrieval models using
% the question-derived querylog. The Stack Exchange dump contains over $5M$
% posts of the type answer.
% We extracted $5M$ posts of the type \textit{Answer} to
% the above $5.5M$ questions.
%% TODO: does the following sentence belong here?
%rishi: no
% Several questions in CQA forums do not have answers, but we retain
% the queries extracted from these questions as they enrich the user profiles.

\subsubsection{Acceptance and votes to {\em qrels}}
\label{subsec:qrels}

% Certain structural features of CQA forums encode relevance judgments.
% First of all, an author of an answer posted as a response to a question
% considered the answer relevant to the question.
% The quality of such annotation can be further estimated using
% the up- and down-votes,
% which can be thought of as community-crowdsourced judgments
% of answer relevance.
% Esp. in moderated communities such as StEx -
% people are discouraged/prevented from posting garbage.

Most popular online CQA forums now have two features that
express relevance: (i) answer acceptance, and (ii) upvotes and downvotes. 
Acceptance is marked by the user who asked the question when an answer
satisfies her information need, and hence can be considered
a gold relevance judgment~\cite{bailey:08} (up to one per question). 
Upvotes and downvotes are generally given by people who understand
the discussion, and hence can be considered silver annotations
(domain experts in~\cite{bailey:08}). 
Gold and silver judgments are more useful than bronze judgments,
which are obtained from annotators who neither issued the query,
nor are experts on the topic. 
% Due to resource constraints, one usually has to make do with
% bronze-quality judges (say, through crowdsourcing). 
% Fortunately, CQA forums give us the means to have a good quality testbed
% with more authentic and credible relevance assessments (or \textit{qrels}
% in TREC parlance). 
% In our resource, we provide the answers that were given in response
% to a question, and the one which was finally accepted. 

For the purpose of this analysis, we use a three-point graded relevance.
We assume the answer accepted by the question author to be
\textit{completely relevant} (2), the other answers posted as a response
to the question as \textit{partially relevant} (1), 
% and answers to other questions in the same topic (subforum) as
% \textit{non-relevant} (0). 
and any other answer as \textit{non-relevant} (0). 
While the assumption about the non-relevance is imperfect,
the concern is alleviated by the fact that Stack Exchange is
a highly moderated forum with questions often closed or marked as
duplicates by the moderators.
It is also worth noting that answers from the same subforum but for
a different question are likely to have high word-level overlap with
the original question, 
which might be a reasonable approach to generating quality negative
training examples for IR models.
% Our Stack Exchange data provides us with about $1.5M$ questions
% with an accepted answer. 
Using upvotes and downvotes to deduce further levels of relevance
is a topic of future work.

% On top of such crowdsourced annotations, forums which have
% the mechanism of \emph{Accepted Answers}, including Stack Exchange,
% provide us with the relevance information collected directly from
% the querying user. When an answer is accepted by a question's author,
% it means the answer has satisfied the information need expressed in
% the question. Thus, such information serves as a golden relevance
% ground truth, and can be automatically mined at scale from
% the source dataset. The Stack Exchange dump provides us with
% X questions with accepted answers, and Y questions contain at least
% one response.

% We derive two scale relevance information [...]
% Otherwise - other answers can be considered relevant, with
% the relevance proportional to the number of upvotes.

	% !TEX root = ../2019-arxiv-stacklog.tex
\subsection{Empirical analysis of the collection}
\label{sec:stats}

In this subsection, we examine the characteristics of a collection derived 
using the methodology described in Sec.~\ref{sec:method}
to shed light on its usefulness. 
% We release the derived collection for further research.

% \subsection{Stack Exchange source data} already covered in Sec. 2
%In this section, we report detailed statistics about our collection that
% show its size and diversity. 
%We also show that it can significantly discriminate between personalized
% ranking algorithms and leave substantial room to show improvement.

% We first provide some descriptive statistics about StackLog,
% and follow them up with simple retrieval experiments.

\subsubsection{Corpus statistics}
\label{subsec:corpus-stats}

\begin{figure*}[t]
	\centering
	\includegraphics[width=\textwidth]{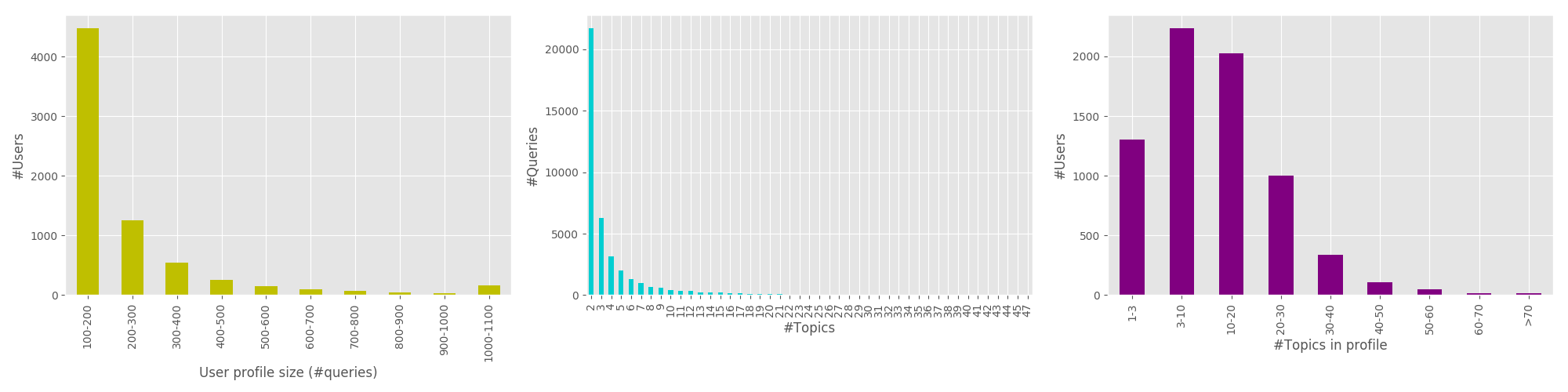}
	%\medskip
	%
	%\includegraphics[width=.3\textwidth]{query-doc-dist.png}\quad
	%\includegraphics[width=.3\textwidth]{topic-query-dist.png}\quad
	%\includegraphics[width=.3\textwidth]{query-doc-dist.png}\quad
	\caption{Distributions describing key features of the presented resource.}
	\label{fig:dists}
\end{figure*}

\begin{table} [t]
	\centering
	%\newcolumntype{G}{>{\columncolor [gray] {0.90}}c}
	%\resizebox{\columnwidth}{!}{
		\begin{tabular}{l r}
			%\begin{tabular}{c G G c c}
			\toprule
			%\textbf{Property}								& \textbf{Value}		\\ \toprule
			No. of users									& $7,104$					\\ 
			No. of queries (distinct) 						& $1,931,336 \; (1,036,953)$	\\ 
			No. of documents								& $5,262,125$				\\
%			No. of query-document pairs with \textit{qrels}	& $5,166,364$ 			\\
			No. of topics (forums)							& $152$					\\ \midrule
			Mean (SD) query length in words					& $2.45 \; (1.74)$ 			\\
			Mean (SD) document length in words				& $126.26 \; (149.22)$		\\
			Mean (SD) user profile size in queries			& $271.87 \; (1392.59)$		\\
			Mean (SD) topics in user profile				& $12.99 \; (10.88)$			\\ \bottomrule
% 			Mean (SD) documents per query 					& $20 (10)$				\\
% 			Mean (SD) queries per topic						& $450 (790)$				\\ \bottomrule
		\end{tabular} %}
	\caption{Aggregate statistics for the derived collection.}
	\label{tab:basic}
\end{table}

\begin{table} [t]	
	\resizebox{\columnwidth}{!}{
	\begin{tabular}{p{2.5cm} p{8cm}}
		\toprule
		\textbf{Topic}	&\textbf{Query with excerpt from accepted answer} \\ \toprule
		Fitness			& \textit{\textbf{Query:}} \textit{knee sleeve strengthen} \\
		& \textit{\textbf{Answer}}: I have osteochondritis desicans (sp?) in my left knee. Basically a nerve is pinched when my support muscles tire and sometimes the pain is bad enough that my knee buckles... But what really helped was lots of squats and bar bell training (deadlift, front squat, back squat, various Olympic lifts)...\\ 
		Beer 			& \textit{\textbf{Query:}} \textit{cider pasteurized orchard} \\
		& \textit{\textbf{Answer}}: I haven't ever been able to tell the difference between a pasteurized or non as a juice base in the final product... THAT stuff still can make a good final beverage, but it turns out quite a bit different than the fresh-pressed stuff.  In particular, the store juice tends to be much more tart...	 \\
		Parenting		& \textit{\textbf{Query:}} \textit{snack children sleep} \\
		& \textit{\textbf{Answer}}: Does a late evening snack improve the odds that a child will sleep well?... But it seems more likely that bedtime snack has become a ritual needed that helps relax a child...	\\ %\midrule
		Programming		& \textit{\textbf{Query:}} \textit{pgp gpl licensed code} \\ 
		& \textit{\textbf{Answer}}: The GPL says you are to free to run, distribute, modify and study the library for any purpose. It does not restrict commercial usage of software...]	\\ 
		Engineering		& \textit{\textbf{Query:}} \textit{torsional fem derivation constant} \\
		& \textit{\textbf{Answer}}: You can find an implementation of a finite element used in computation of arbitrary shape section torsional constant here...	\\ \bottomrule
	\end{tabular}}
	\caption{Examples from the derived collection.}
	\label{tab:examples}
\end{table}

% We excluded the largest subforum Stack Overflow from the source dataset
% to avoid the dominance of programming queries in our collection. 
% We also excluded the users for whom we were unable to extract
% \textit{at least $100$ queries}, so that each user in StackLog
% has enough content in her log to enable individualized modeling. 
% The resulting filtered data constitutes StackLog.

% StackLog is available for download under the following anonymous
% Dropbox link:
% \url{https://www.dropbox.com/s/xawrv8naze7pmkq/StackLogQueries.tsv.zip}
% (Accessed 28 Feb 2017).
% Each line in the query file consists of three fields separated by tabs:
% user id, post id (in the format \emph{subforum - post id - timestamp}),
% and the query. 
% The original questions, documents and relevance judgments are also present
% in the archive.

Table~\ref{tab:basic} and Fig.~\ref{fig:dists} present basic statistics over
the elements of the collection.
There are about $7K$ user profiles, together issuing approximately $2M$ queries. 
The distribution of user profile sizes is shown in Fig.~\ref{fig:dists} (a). 
%Fig.~\ref{fig:dists} (b) shows the query length distributions, computed
% over all queries. 
%As discussed in Sec.~\ref{sec:method}, these distributions follow
% the ones observed in the 2006 AOL log~\cite{pass2006picture}.

The associated document collection size is about $5M$, created from
all answer posts. $1.5M$ of these are an accepted answer to one of
the queries, and hence are \textit{completely relevant} for those queries. 
The rest of the documents ($\simeq 3.5M$) are \textit{partially relevant}
for those queries to which they were posted as answers. 
% Hence, we can claim that each of the $5M$ documents has
% an associated relevance judgment. 
% This number is far higher than in other IR benchmarks. 
The average document consists of about 126 words, with a standard deviation
of about $149$ words. 
% So our documents are indeed usually a few paragraphs long and not
% trivial snippets of one or two sentences.

User profiles display rich topical variety, with the average number of
distinct subforum contributions from a user being about $13$, with
a standard deviation of about $10$.  
The user-subforum distribution is shown in Fig.~\ref{fig:dists} (c). 
While this diversity in forum contributions suggests that there is a scope
for topical personalization,
another important aspect to personalization is topical ambiguity.
Fig.~\ref{fig:dists} (b) shows the distribution of the number of queries
($y$-axis) which are textually equal, but were derived from a number of
different topics ($x$-axis). We removed the bar for queries derived from
a single topic only (there were around $1M$ such queries).
Table~\ref{tab:examples} presents representative queries from the collection
together
with excerpts from the accepted answer document (gold relevance).

\subsubsection{Performance in retrieval}
\label{subsec:ranking}

\begin{table} [t]
	\centering
%	\caption{Performance of basic retrieval methods on the derived collection.}
	%\newcolumntype{G}{>{\columncolor [gray] {0.90}}c}
	%\resizebox{\columnwidth}{!}{
		\begin{tabular}{l c c}
			%\begin{tabular}{c G G c c}
			\toprule
			\textbf{Method}			& \textbf{MAP}	& \textbf{MRR} 	\\ \toprule
			Indri					& $0.076$		& $0.053$		\\ 
% 			Indri + Re-rank			& $0.089$		& $0.061$		\\                      \midrule
			Indri + q2a				& $0.398$	 	& $0.211$		\\ \bottomrule
% 			Indri + Re-rank + q2a	& $0.403$		& $0.212$		\\ \bottomrule
		\end{tabular} %}
		% \\ * denotes significant differences under the $2$-tailed paired $t$-test ($p < 0.05$).
		% rishi: do we need this remark? are we trying to prove something?
	\caption{Basic retrieval performance.}
	\label{tab:ranker}
\end{table}

We perform basic retrieval experiments to gain insights into
the retrieval difficulty of the collection.
First, we index the documents with Indri~\cite{strohman2005indri},
using the standard stopword list (\url{https://www.ranks.nl/stopwords})
and the Porter Stemmer~\cite{porter2001snowball}.
We then retrieve the top-$100$ documents for each of the $2M$ queries in
the query log,
using Dirichlet smoothing~\cite{zhai2017study}.
%with parameter $\mu$ set to the average document length. 

Effectiveness is assessed using:
(i) \emph{Mean Average Precision (MAP)}~\cite{manning2008introduction}
is computed considering the documents
which originally were the answers to corresponding questions as relevant;
and (ii) \emph{Mean Reciprocal Rank (MRR)}~\cite{voorhees1999trec} is computed
using
the answer accepted by
the asker as relevant.
All measures are averaged over queries for each user, and then macro-averaged
over users.

% Table~\ref{tab:ranker} presents the performance of four retrieval baselines.
Table~\ref{tab:ranker} presents the performance of two retrieval baselines.
The \emph{Indri} method, representing the raw Indri retrieval, leaves a lot
of room for improvement.
There are two main reasons for this. First, when long questions are reduced
to very short queries, 
often a large pool of documents match the query, possibly leaving
the relevant documents beyond the top-$100$ results.
Second, since the document collection consists of posts of
type \textit{answer}, the vocabulary of the questions need not match
the document literally.
%Second, since the document collection consists of posts of type answer,
% the vocabulary of the questions occur in the answer's context only implicitly. 
%To understand the collection behavior better when the vocabulary contexts
% are explicit, 
To better understand this vocabulary mismatch issue in the collection,
we perform retrieval with Indri over a collection where the questions are 
appended to the answers to form the documents.
%propagated to answers (i.e., we append the source question to each of
% the answers). 
The \emph{Indri+q2a} row in Table~\ref{tab:ranker} quantifies to what extent
this influences performance.
%the contexts only being implicit have influenced the raw
% retrieval performance in the $Indri$ method.

% To investigate the leeway for personalization, we implement
% a basic personalized re-ranker, which operates on the top-$k$ results
% returned by Indri. 
% Let $\Phi_u$ be the distribution of a user $u$'s queries across
% the subforums. Each document $d$ in the ranking has its subforum
% of origin $s$. 
% We compute the personalized ranking for user $u$ by recomputing
% the score of each of the top-$100$ documents $d$ as follows:
% \vspace{-10px}
% \begin{equation}
% 	score_u(q, d) = score_{Indri}^{norm}(q, d) \cdot \Phi_u(s)
% \end{equation}
% where $score_{Indri}^{norm}(q, d) = score_{Indri}(q, d) + \min_{d*} score_{Indri}(q, d*)$, and $s$ is the subforum where the document comes from. 
% The two rows in Table~\ref{tab:ranker} with \textit{Re-rank} show that this simple re-ranking heuristic improves the IR performance on the StackLog collection. 
\noindent \textbf{Summary.} Results suggest that the derived collection
leaves ample room for improvement
% to
for more
% advance
advanced methods.
\section{Related work}
\label{sec:related}

\if{0}
\textbf{Collections for personalization studies.} Previous studies on 
personalization either had access to commercial Web search 
data~\cite{shokouhi:15,singla:14,wang:13} or used proxy methods. For example, 
Sieg et al.~\cite{sieg:07} used %\textit{automatically generated queries}
automatically generated queries
%rishi: no need to emphasize, diminishes novelty
from specific categories of documents to evaluate their personalization 
strategy. Hannak et al.~\cite{hannak:13} collect popular queries from Google 
Zeitgeist and WebMD, create new Google accounts by user simulation, and pay AMT 
workers with existing Google accounts to test the effect of personalization.
Tan 
et al.~\cite{tan:06} asked four volunteer computer science students to use a 
plug-in that logged their Google queries and the top-20 retrieved documents, 
which were later assessed by the volunteers. Limitations of the TREC Sessions 
data~\cite{carterette:14} and the Yandex anonymized corpus~\cite{serdyukov:14} 
have been discussed in Sec.~\ref{sec:intro}.
% The 2014 Yandex collection released as part of the Log-based Personalization 
%Workshop WSCD 2014~\cite{serdyukov:14} is a very useful resource for evaluating 
%personalization algorithms, and was used in some research 
%studies~\cite{ustinovskiy:15,cai:14}. However, for privacy issues, every query 
%word or URL is replaced by a meaningless numeric 
% %IDs\footnote{\url{
%https://www.kaggle.com/c/yandex-personalized-web-search-challenge/data}, 
%Accessed 27 February 2017.}. This makes semantic interpretability
% impossible and 
%may be a reason why this resource did not become very popular.
% The \textit{TREC 
%Sessions Track 2014 data}~\cite{carterette:14} has $148$ users, $\simeq 4.5k$ 
%queries and $36k$ retrieved documents. There are roughly ten sessions per user, 
%where each session
\fi

\noindent \textbf{Collections for privacy studies.} Research on 
privacy-preserving search has been perennially plagued by a scarcity of 
corpora with interaction and profile information. 
Consequentially, quite a few works on privacy remain theoretical or proposals 
without empirical validation~\cite{saint:07}. 
Otherwise, it is a common practice to resort to using
the 2006 AOL logs~\cite{pass2006picture}
despite the controversial circumstances of its 
release~\cite{biega2016r,shou:14}.
% despite the ethical issues~\cite{biega2016r,shou:14}. 
Volunteers have shared their search profiles in exceptional cases~\cite{xu:07},
but this may lead to a feeling of regret later on.

% \noindent \textbf{Collections with topical queries?} TREC data like medical 
%and legal? See \url{http://trec.nist.gov/data.html} 

% qru-1 QRU-1~\cite{li:12}, a small artifical query log with XYZ queries. No 
%user information. QRU-1 has nothing other than queries. Not worth citing.

\noindent \textbf{CQA datasets.} The idea of tapping into CQA for
curated resource creation has been 
around % in the prior literature.
for a while.
For example, there exists a small collection of crowdsourced queries based on 
questions from Yahoo! Answers~\cite{habernal2016new}. However, to the best of 
our 
knowledge, no large scale query collections with detailed user histories and 
relevance judgments have been extracted from CQA datasets. Harnessing CQA 
resources is an active topic now, and datasets like
duplicate questions (\url{https://bit.ly/2upwz0x})
and question-code pairs~\cite{yao2018staqc}
have recently been extracted. However, these resources are not suitable for
directions discussed here.

\noindent{\textbf{Reducing queries.}} With regard to methodology, our work is 
related to \textit{verbose query reduction}
~\cite{gupta2015information,huston2010evaluating,arguello2017using}.
% Converting questions to queries can also be seen as the problem of 
%\textit{verbose query reduction}. 
A number of such techniques have been evaluated in the context of CQA 
forums~\cite{huston2010evaluating}. 
% Another study
% rishi: doesn't sound good
Kumaran and Carvalho~\cite{kumaran2009reducing} looked at
query reduction based on query quality 
predictors, including IDF-based scores.
We note, however, that these techniques aim at producing short queries that 
maximize retrieval effectiveness,
while our goal is to produce queries that resemble those issued by real users.
Since we also release the original NL questions along with the queries, the 
community is encouraged to explore other query reduction 
techniques~\cite{balasubramanian:10,park:10} which can contribute to an improved 
version of our % the
% StackLog
resource.

	% !TEX root = ../2019-arxiv-stacklog.tex
\section{Conclusion}
\label{sec:conext}

In this paper, we conducted a user study to provide a better understanding
of how humans formulate queries
from information needs described by natural language questions.
Gaining insights into this process forms an important foundation
for automatic conversion methods, which would allow us
to create IR collections from the publicly available CQA resources,
Such collections with rich user profiles,
unavailable to academic researchers otherwise,
would be a great asset driving research on user-centric privacy.

Beyond query log synthesis, our paper proposed a methodology
for deriving other IR collection elements from the data
and structure of CQA forums, including documents
and gold relevance judgments. We further empirically analyzed
a collection derived from Stack Exchange,
showing its potential as a difficult retrieval benchmark.
We release a dataset of $7,000$ crowdsourced question-query pairs
as well as the derived collection to foster further research
in automatic derivation of large-scale privacy-friendly
IR collections from CQA forums.

% We released % StackLog --
% a query log derived from the Stack Exchange CQA forum. 
% Along with queries, documents and gold relevance judgments,
% it also includes user-ids and topics --
% making it ideal for research on personalization and privacy. 
% As seen from low MAP values achieved by baseline algorithms,
% this collection is expected to be a difficult and
% realistic benchmark
% % , leaving significant leeway
% for more advanced techniques. 
% Further, the general methodology for constructing the dataset
% can be harnessed for creating similar resources from CQA forums.
% 
% Since the source data is publicly available, there are
% multiple ways in which this resource can be extended
% by the community. 
% Smarter ways of deriving a query from a question,% a question to a query,
% estimating additional relevance from upvotes and downvotes,
% and simulating search sessions from time-stamped ``related questions''
% are only a few of several such promising future directions.

	\bibliographystyle{splncs04}
	\bibliography{qa_qlog}
	
\end{document}